\documentstyle [equations,twoside,12pt]{report}
\newcommand{\be}{\begin{equation}}
\newtheorem{prop}{Proposition}
\newcommand{\ee}{\end{equation}}
\newcommand{\bea}{\begin{eqnarray}}
\newcommand{\eea}{\end{eqnarray}}
\newcommand{\bes}{\begin{subequations}} 
\newcommand{\beq}{\begin{eqalignno}}
\newcommand{\ees}{\end{subequations}} 
\newcommand{\eeq}{\end{eqalignno}}
\newcommand{\lab}{\label}
\newcommand{\half}{\frac{1}{2}}

\def\Journal#1#2#3#4{{#1} {\bf #2}, #3 (#4)}

\def\JMP{\em J. Math. Phys.}

\def\PLB{{\em Phys. Lett.}  B}

\def\PRD{{\em Phys. Rev.} D}

\def\ra{\rightarrow}

\def\al{\alpha}

\newcommand{\ga}{\gamma}
\newcommand{\Ga}{\Gamma}
\newcommand{\ep}{\epsilon}
\def\lab{\label}

\begin{document}
\setlength{\baselineskip}{.7cm}
$$ $$

\begin{center}
\huge{A class of nonlinear wave equations containing the continuous
  Toda case} 
\end{center}
\begin{center}
\large{E. Alfinito$^{c,b}$, M.S. Causo$^{a,b}$, G. Profilo$^{d}$, and
  G.  
Soliani$^{a,b}$}. 
 \\ 
\vspace{0.2cm}
\normalsize
$^{a}${\it Dipartimento di Fisica dell'Universit\`a, Lecce, Italy}\\
$^{b}${\it  Istituto Nazionale di Fisica Nucleare, Sezione di Lecce,
Italy},\\
 $^{c}${\it Dipartimento di Fisica dell'Universit\`a, Salerno, Italy},\\
  
 $^{d}${\it Dipartimento di Matematica dell'Universit\`a, Lecce,
 Italy.}
\end{center}

\vspace{1.5cm}

\centerline{\bf Abstract}
\vspace{0.2cm}
\small
We consider a nonlinear field equation which can be derived from a 
binomial lattice as a continuous limit. This equation, containing
a perturbative friction-like term and a free parameter $\gamma$,
reproduces the Toda case (in absence of the friction-like term) and other
equations of physical interest, by choosing particular values of
$\gamma$. We apply the symmetry and the approximate symmetry
approach, and the prolongation technique. Our main purpose is 
to check the limits of validity of different analytical methods in
the study of nonlinear field equations. We show that the equation
under investigation with the friction-like term is characterized
by a finite-dimensional Lie algebra admitting a realization in terms
of boson annhilation and creation operators. In absence of the
friction-like term, the equation is linearized and connected with
equations of the Bessel type. Examples of exact solutions are 
displayed, and the algebraic structure of the equation is discussed.
\normalsize

\newpage 
\section*{1 Introduction}

In this paper we study the class of nonlinear wave equations
\be 
u_{tt}\,+\,\epsilon u_{t}\,=\,\left[\left(1+\frac{u}{\gamma}
\right)^{\gamma -1}\right]_{xx},
\lab{1.1}\ee

where $u=u(x,t)$, subscripts denote partial derivatives, $\epsilon$ is a
free 
parameter, and $\gamma$ is a constant which takes all (real) values
except zero 
and one.

For $\epsilon=0$, Eq.(\ref{1.1}) can be considered as the continuous
limit of a 
uniform one-dimensional nonlinear lattice of $N$ particles interacting
through 
the nearest-neighbour potential \cite{LLS} 
\be
\phi(r_{n})\,=\,\frac{a_{n}}{b_{n}}\left[\left(1+\frac{b_{n}\ r_{n}}{\ga}
\right)^{\gamma}-\left(1+ b_{n}\ r_{n}
\right)\right],
\lab{1.2}\ee 
where $a_{n}$ and $b_{n}$ are constants of the $n$-th nonlinear spring.

The equation of motion of the chain reads
\be
M_{n}\,{\ddot y}_{n}=-\left[\phi^{\prime}(r_{n+1})
-\phi^{\prime}(r_{n})\right], \lab{1.3}\ee 
where $\phi^{\prime}=\frac{d\phi}{dr}$, $M_{n}$ is the mass of the
$n$-particle and $r_{n}=y_{n}-y_{n+1}$, $y_{n}$ being the displacement
of the 
$n$-particle from its the equilibrium position. We shall call ``binomial
lattice" the 
chain governed by the potential (\ref{1.2}). If the binomial lattice is
uniform, 
i.e. $a_{n}=a$, $b_{n}=b$ and $M_{n}=M$ for any $n$, the equation of
motion  
(\ref{1.3}) takes the form
\be
{\ddot r}_{n}=
\frac{a}{M}\left[\left(1+\frac{b}{\ga}\ r_{n+2}\right)^{\ga -1}
-2\left(1+\frac{b}{\ga}\ r_{n+1}\right)^{\ga -1}+\left(1+\frac{b}{\ga} \ r_{n}
\right)^{\ga -1}\right].
\lab{1.4}\ee 

For $\ga \rightarrow \infty$, $\ga=2$, $\ga=3$ and $\ga=-1$, Eq.(\ref{1.4})
reproduces the Toda \cite{Toda}, the harmonic, the 
Fermi-Pasta-Ulam \cite{FPU}
 and a Coulomb-like uniform lattice, respectively. 
Since Eq.(\ref{1.4}) models
interesting physical situations, its analysis may be important mostly 
from the
point of view of a unifying theory of lattice systems described by the
potential (\ref{1.2}).

This difficult task is made simpler by starting from the 
investigation of the
continuous limit of Eq.(\ref{1.4}). 
Then, Eq.(\ref{1.4}) becomes 
\be
u_{tt}\,=\,\left[\left(1+\frac{u}{\gamma}
\right)^{\gamma -1}\right]_{xx},
\lab{1.5}\ee
after the rescaling $b\,r\,\rightarrow u$ and $\sqrt{\frac{ab}{M}} \ t 
\rightarrow t$.

For $\gamma \rightarrow \infty$, Eq.(\ref{1.5}) coincides with the 
(1+1)-dimensional continuous
Toda equation \cite{Saveliev}
\be
u_{tt} = \left(e^{u}\right)_{xx},
\lab{1.6}\ee
which is known also as {\em heavenly equation} and plays a crucial 
role in General
Relativity \cite{Finley}, \cite{Boyer}.

Equation ({\ref{1.1}}) represents a modified version of Eq.(\ref{1.5}). It
contains a friction-like term which can be considered as a small 
perturbation.
This extension of Eq.({1.5}) allows us to check the validity of different
analytical methods usually employed to handle  integrable equations.

Precisely, here we apply to Eq.(\ref{1.1}) the 
symmetry \cite{Olver} and the approximate symmetry approach 
\cite{Ibragimov}, and 
the prolongation
procedure \cite{Wal}.

The main results achieved in this paper are the following. In Sec. 2
 we find the approximate symmetry generators of Eq.(\ref{1.1}) in 
the case in which $\epsilon$ is a perturbative parameter.
The commutation relations among these operators do not define an
{\em exact} finite-dimensional algebra. (Exact algebras arise only for
$\epsilon=0$ and when $\epsilon \ne 0$ is not considered as a small
quantity).
Notwithstanding, an exact finite-dimensional Lie algebra can be
constructed by introducing certain auxiliary operators.
A realization of this algebra, which characterizes the approximate
 symmetries of Eq.~(\ref{1.1}), is obtained in terms of boson
annihilation and creation operators.

In Sec. 3, the symmetry approach is exploited to determine special
solutions of Eq.(\ref{1.1}) and Eq.(1.5). In this context, the study
of the symmetry corresponding to the generator ${X^0}_4$ (see (2.5c))
 is of particular interest. In fact, ${X^0}_4$ is a scale symmetry
in the variables $x, u$ together with a translation of $\frac{2\gamma}
{\gamma -2}$ along $u$.
For $\gamma \to \infty$, i.e. when Eq.(1.1) coincides with the
continuous Toda equation, the scale symmetry is broken.
The reduced equation related to ${X^0}_4$ resembles the 
one-dimensional Liouville equation (see (3.4)) containing a square
derivative which vanishes in the limit $\gamma \to \infty$.
In the case $\epsilon = 0$, Eq.(3.4) is exactly solved for some
values of the parameter $\gamma$. For example, if $\gamma=3$ (a 
choice associated with the Fermi-Pasta-Ulam potential), we get a solution
to Eq.(1.5) via Eq.(3.4) expressed in terms of the Weierstrass function.

For $\epsilon \ne 0$, we provide an implicit solution to 
Eq.(1.1) by using the symmetry variable $V=v\partial_x + \partial_t$,
 where $v$ is a constant. This solution (see (3.42)) involves a
 generalization of the exponential-integral function. It is notable
 that for $\gamma \to \infty$, this solution reproduces just that corresponding 
 to the continuous Toda case \cite{Ele}.
 
 Sec. 4 is devoted to a study of Eq.(1.1) within the prolongation scheme 
\cite{Wal}.
 This method is particularly convenient because it allows us to
 analyze Eq.(1.1) for $\epsilon=0$ and $\epsilon \ne 0$ by an algebraic
 point of view. We show that the (differential) 
 prolongation equations for Eq.(1.5)
 afford a class of solutions connected with the Lie algebra of
 the  Euclidean
 group $E_2$ in the plane. 
 This algebra leads to a linearized version of Eq.(1.5), in the sense
 that a mapping is established between Eq.(1.5) and the linear
 wave equation (4.6) written in a pseudopotential variable.
For $\gamma \to \infty$, Eq.(4.6) coincides with a two-dimensional
 form of a linear wave equation that occurs in a generalized 
Gibbson-Hawking ansatz ~\cite{Lebrun}. On the other hand, the 
prolongation equations for Eq.(1.1) can be solved in terms
of power series expansions whose coefficients (in the variable
$z=\left(1+\frac{u}{\gamma}\right)^{\gamma-1}$ ) depend on the
pseudopotential components and, for a certain infinite set
of values of $\gamma$, obey a finite-dimensional Lie algebra.
A representation of this algebra indicates the existence of a 
possible link between the prolongation method and the
symmetry approach. Finally, in Sec. 5, some comments are reperted.

\setcounter{chapter}{2}
\setcounter{equation}{0}
\section*{2 The approximate symmetry approach}

\par In the case in which $\epsilon$ is a perturbative parameter,
Eq.(1.1) can be handled profitably by means of the method 
devised by Baikov, Gazizov and Ibragimov some years ago \cite{Ibragimov}.
This method enables one to construct approximate symmetries
which are stable for small perturbations of the differential 
equation under investigation.

In order to obtain the approximate symmetries of Eq. (1.1) ($\epsilon 
\neq 0$), we have to introduce the quantities
\bea
F_{0} & = & u_{tt} -\left[\left(1+u/\gamma\right)^{\gamma-1}\right]
_{xx},\nonumber \\
F_{1} & = & u_{t},\lab{Y.1}
\eea
into the equations (3.16) and (3.17) of \cite{Ibragimov}.

For brevity, we shall omit any calculation and report below the results
only.

The approximate symmetry generator turns out to be
\bea
X  &=&  \left[c_{1}t+c_{2}
 +\epsilon \left(\frac{\gamma-2}{\gamma+2}c_{1}
\frac{t^{2}}{2}+
k_{1}t+k_{2}\right)\right]\partial_{t}
+\left[ \left( c_{1}+c_{2}\right)x+c_{4}
+\epsilon \left(\left(k_{1}+k_{3}\right)x
 +k_{4}\right)\right]
\partial_{x} \nonumber \\
& & +\left[\frac{2\gamma}{\gamma-2}\,c_{3}
\left(1+\frac{u}{\gamma}\right)+2\epsilon\frac{\gamma}{\gamma-2}
\left(k_{3}-
\frac{\gamma-2}{\gamma+2}c_{1}t\right)
 \left(1+\frac{u}{\gamma}\right)\right]\partial_{u},\lab{Y.2}
\eea
\par with $c_{1}$, $c_{2}$,... and $k_{1}$, $k_{2}$,...
 arbitrary constants.

From (\ref{Y.2}) we get
\be
X_{1}={X_{1}}^{0}+\epsilon\left[\frac{\gamma-2}{\gamma+2}
\frac{t^{2}}{2}\partial_{t}
-\frac{2\gamma t}{\gamma+2}\left
(1+\frac{u}{\gamma}\right)\partial_{u}\right],\lab{Y.3} \\
\ee
where
\be
X^{0}_{1}=t\partial_{t}+x\partial_{x}, \lab{Y.4}
\ee
\bes\lab{Y.5}
\beq 
X_{2} & \equiv \, {X_{2}}^{0}=\partial_{t},\label{Y.5-a}\\
X_{3} & \equiv \, {X_{3}}^{0}=\partial_{x},\label{Y.5-b} \\
X_{4} & \equiv \, {X_{4}}^{0}=x\partial_{x}+\frac{2\gamma}
{\gamma-2}\left(1+\frac{u}{\gamma}\right)\partial_{u},
\label{Y.5-c} \\
X_{5} & = \, \epsilon \left(t\partial_{t}+x\partial_{x} \right),
\label{Y.5-d} \\
X_{6} & = \, \epsilon \partial_{t},\label{Y.5-e} \\
X_{7} & = \, \epsilon \left[x\partial_{x} + \frac{2\gamma}{\gamma - 2}
 \left(1+\frac{u}{\gamma}\right)\partial_{u}\right],\label{Y.5-f} \\
X_{8} & = \, \epsilon \partial_{x}.\label{Y.5-g} 
 \eeq
\ees

The operators $X^{0}_{1}$, $X^{0}_{2}$, $X^{0}_{3}$, $X^{0}_{4}$ 
are the (exact) symmetry generators of Eq. (1.1) for $\epsilon=0$, 
while $X_{1}$, $X_{2}$, $X_{3}$, $X_{4}$ are the {\it approximate}
symmetry 
generators of Eq. (1.1) for $ \epsilon \ne 0$. The {\it exact} symmetry
 generators of Eq.(1.1) for $\epsilon \ne 0$ are $X^{0}_{1}$,
$X^{0}_{2}$ and $X^{0}_{3}$. The operators $X_{5}^{0},\;X_{6}^{0},\;
X_{7}^{0},\;X_{8}^{0}$ are inessential.

The approximate symmetry generators $X_{1}$, $X_{2}$, $X_{3}$, $X_{4}$
satisfy the commutation relations
\bea
 \left[X_{2},X_{1}\right] &  = & X_{2}
+ \epsilon\left[ \frac{\gamma - 2}{\gamma + 2}
\,t\,\partial_{t}
-\frac{2\gamma}{\gamma+2}\left(1+\frac{u}{\gamma}\right)
\partial_{u}\right],\lab{Y.6} \\
\quad \left[X_{2},X_{4}\right] & = & 0,
\quad\left[X_{2},X_{3}\right]  =  0,
\quad\left[X_{1},X_{4}\right] = 0,\lab{Y.7} \\
\quad\left[X_{1},X_{3}\right] & = & -X_{3},
\quad\left[X_{3},X_{4}\right] = X_{3}\lab{Y.8} 
\eea

The commutation rules (\ref{Y.6})-(\ref{Y.8}) 
define an exact finite-dimensional Lie
 algebra only for $\epsilon = 0 $. However, they can be exploited to
build up a realization of an exact Lie algebra, which holds for 
$\epsilon \ne  0 $, with the help of the "auxiliary" operators

\bes\lab{Y.9}
\beq 
Z & =  \frac{\gamma - 2 }{\gamma + 2 } t \partial_{t}-
\frac{2\gamma}{\gamma+2}\left(1+\frac{u}{\gamma}\right)\partial_{u},
\lab{Y.9-a}\\
Y & = \frac{\gamma-2}{\gamma+2}\frac{t^{2}}{2}\partial_{t}-
\frac{2\gamma t}{\gamma+2}\left(1+\frac{u}{\gamma}\right)\partial_{u}.
\lab{Y.9-b}
 \eeq
\ees


In doing so, we find
\bes\lab{Y.10} 
\beq
 \left[X_{2},X_{1}\right] & = \, X_{2} +\epsilon Z, \lab{Y.10-a} \\
\left[X_{2},X_{4}\right] & =\,  \left[X_{2},X_{3}\right]
=\left[X_{1},X_{4}\right]=0\lab{Y.10-b} ,\\
\left[X_{1},X_{3}\right] & = \, -X_{3},
\quad\left [X_{3},X_{4}\right]=X_{3},\lab{Y.10-c} \\
\left[X_{1},Z\right] & = \, -\epsilon \frac{\gamma-2}{\gamma+2}Y,
\quad \left[X_{1},Y\right]=Y,\lab{Y.10-d}\\
\left[X_{2},Y\right] & = \, Z,\quad \left[X_{2},Z\right]
=\frac{\gamma-2}{\gamma+2}X_{2},
\quad \left[Z,Y\right]=\frac{\gamma-2}{\gamma+2}Y,\lab{Y.10-e}\\
\quad\left[X_{3},Y\right] & = \, \left[X_{3},Z\right]
=\left[X_{4},Y\right]=\left[X_{4},Z\right]=0.\lab{Y.10-f}
\eeq
\ees

Hereafter, the symbols $Y$, $Z$, $X_{j}$ ($j=1,..,4$) will mean both 
the abstract elements and the corresponding realizations (see (\ref{Y.5}),
(\ref{Y.9}) and (\ref{Y.10})) of 
the finite-dimensional Lie algebra (\ref{Y.10}).

At this stage, we remark that the subalgebra (\ref{Y.10-e}) 
is isomorphic to the
$sl(2, R)$ algebra
\be
 \left[Z^{\prime},T\right]=2S,\quad \left[T,S\right]
=2Z^{\prime},\quad
\left[S,Z^{\prime}\right]=-2T,\lab{Y.11}
\ee
where
\bea
T & = & \sqrt{2\frac{\gamma+2}{\gamma-2}}\ \left(Y+X_{2}\right),
\qquad S=\sqrt{2\frac{\gamma+2}{\gamma-2}}\ \left(Y-X_{2}\right),\nonumber\\
Z^{\prime} & = & 2\frac{\gamma+2}{\gamma-2}\ Z .\lab{Y.12}
\eea

It can be proved straightforwardly that

\begin{prop}
The Casimir operator
\bea
C & = & T^{2}-S^{2}-{Z^{\prime}}^{2}\nonumber\\
  & \equiv & 4 \ \frac{\gamma+2}{\gamma-2}\left[2X_{2}Y-
 Z\left(\frac{\gamma+2}{\gamma-2}Z+1\right)\right] \lab{Y.13}
\eea
of the Lie algebra (\ref{Y.11}), commutes with all 
the elements $Y, Z, X_{j} $
( $j=1,..,4$) of the Lie algebra (\ref{Y.10}).
\end{prop}

Furthermore, the following Proposition holds:
\begin{prop}
The Lie algebra (\ref{Y.10}) admits a realization 
in terms of boson annihilation and creation operators.
\end{prop}

This can be seen by setting

\be
a^{+}_{1}=t,\quad a^{+}_{2}=u, \quad a^{+}_{3}=x,
\quad a_{1}=\partial_{t},\quad a_{2}=\partial_{u},
\quad a_{3}=\partial_{x}, \lab{Y.14}
\ee

to give
\bea
 \left[a_{j},a^{+}_{k}\right] & = & \delta_{jk},\quad
 \left[a_{j},a_{k}\right]=0,
\quad \left[a^{+}_{j},a^{+}_{k}\right]=0, \qquad\hbox{($j,k=1,2,3$)}, 
\lab{Y.15} \\
&& \nonumber\\
Y & = & \frac{\gamma-2}{2\left(\gamma+2\right)}{a^{+}_{1}}^{2}a_{1}-
\frac{2\gamma \ a^{+}_{1} }{\gamma+2}\left(1+\frac{1}{\gamma}a^{+}_{2}\right)a_{2},
\nonumber\\
Z & = & \frac{\gamma-2}{\gamma+2} \,  a_{1}^{+}\  a_{1}
-\frac{2\gamma}{\gamma+2}\left(1+\frac{1}{\gamma}\, a_{2}^{+} \right)a_{2},
\nonumber\\
X_{1} & = & a_{1}^{+}\ a_{1}\,+\,a_{3}^{+} \ a_{3}\,+\,\epsilon \ Y,\quad 
X_{2}\,=\,a_{1},\quad X_{3}\,=\,a_{2},\nonumber\\
X_{4} & = & a^{+}_{3}\ a_{3}+\frac{2\gamma}{\gamma-2}\left(1+\frac{1}{\gamma}
a^{+}_{2}\right)a_{2}, \lab{Y.16}
\eea
and
\be
C=\left(\frac{\gamma}{\gamma-2}\right)^{2}C_{\infty}+C\left(\gamma\right),
 \lab{Y.17}\ee

where $C_{\infty}=-8a_{2}\left(2a_{2}+1\right)$ is the
 Casimir invariant relative
to the continuous Toda case $ \gamma  \rightarrow \infty $, and the
operator $C\left(\gamma\right)$, defined by 
\be
C\left(\gamma\right)=-\frac{8a^{+}_{2}a_{2}}{\gamma-2}\left[1+\frac{2\gamma}
{\gamma-2}
\left(2a_{2}+\frac{1}{\gamma}a^{+}_{2}a_{2}\right)\right],
\lab{Y.18} \ee

tends to zero as $ \gamma \rightarrow \infty $.

Although the full role of the closed algebra (\ref{Y.10}) has to be 
better understood, it is noteworthy that the auxiliary operators $Z$ and $Y$
(see  
(\ref{Y.9})) can be interpreted as symmetry variables 
of the equations
\be u_{xx}\,+\,\ep u_{x}
\,-\,\frac{2}{\ga}\left(1+\frac{u}{\ga}\right)^{-2}  
u_{x}^{2}\,=\,-\left(1+\frac{u}{\ga}\right)^{2}
\left[ \left(1+\frac{u}{\ga}\right)^{1-\ga} \right]_{tt}, 
\lab{Y.19}\ee
\be u_{xx}\,+\,\ep u_{x}
\,-\,\frac{3}{2\ga}\left(1+\frac{u}{\ga}\right)^{-1}  
u_{x}^{2}\,=\,-2\left(1+\frac{u}{\ga}\right)^{\frac{3}{2}}
\left[ \left(1+\frac{u}{\ga}\right)^{\frac{1-\ga}{2}} \right]_{tt}, 
\lab{Y.20}\ee
respectively.

Equations (\ref{Y.19}) and (\ref{Y.20}) arise formally from Eq.(1.1) via 
the transformations
\be 
x\,\rightarrow \,t, \qquad u\,\rightarrow \, -\frac{u}{1+\frac{u}{\ga}},
\lab{Y.21}\ee
and 
\be 
x\,\rightarrow \,t, \qquad u\,\rightarrow \, \gamma\left[\left(1+
\frac{u}{\ga}\right)^{-1/2}-1\right].
\lab{Y.22}\ee

From $Z$ we obtain the invariants $x^{\prime}=x$, $\,\eta^{\prime}(x^{\prime}) 
=\eta(x)$, with
\be
\eta(x)\,=\,
\left(1+\frac{u}{\ga}\right)^{\frac{\ga-2}{2}} t,
\lab{Y.23}\ee
which gives the reduced equation
\be 
\eta_{x}^{2}+\frac{2-\ga}{\ga}\eta\eta_{xx}
+\frac{2-\ga}{\ga}\ep\eta\eta_{x}\,=
\,\frac{\ga-1}{\ga}
\lab{Y.24}\ee
from substitution into (\ref{Y.19}). We notice that Eq.(\ref{Y.24}) 
coincides, formally, with Eq.(\ref{X.3}) for $x\rightarrow t$ (see
Sec. 3). Therefore, 
putting $\eta= e^{W}$, Eq.(\ref{Y.24}) becomes Eq.(\ref{X.4}) where $t$ is 
replaced by $x$.

On the other hand, the operator $Y$ yields the invariants $x^{\prime}=x\,$, 
$\,r^{\prime}(x^{\prime})\,=\,r(x)$, with 
\be 
r(x)\,=\,\left(1+\frac{u}{\ga}\right)^{\frac{\ga-2}{2}} t,
\lab{Y.25}\ee
which leads to the ordinary differential equation of the modified Liouville 
type (see (\ref{X.4}))
\be 
v_{xx}\,+\,\ep v_{x}
\,+\,\frac{1}{\ga-2}\ v_{x}^{2}\,=
\,\frac{2}{\ga-2}\ \left[\frac{1}{\ga}-2(1+\ga)\right]e^{-2v},
\lab{Y.26}\ee
where $v=\ln r$.

We point out that in the continuous Toda case, i.e. for $\ga \rightarrow 
\infty$, in the reduced equations of the one-dimensional modified 
Liouville type [(\ref{X.4}),(\ref{X.4}) written in the variable $x$ and 
(\ref{Y.26})], the terms $W_{t}^{2}$, $W_{x}^{2}$, and  $v_{x}^{2}$ 
disappear. This is due to the fact that for $\ga\rightarrow \infty$, the 
coefficient in front of $u \partial_{u}$ in $X^{0}_{4}$ is vanishing, so 
that  $X^{0}_{4}$ is not longer a scale symmetry in the variables 
$x,\;u$ (together with a translation of $\frac{2\ga}{\ga -2}$ along $u$). 
Thus, the presence of square first derivatives in the reduced equations 
generated by  $X^{0}_{4}$ reflects the existence of a scale symmetry, 
which is broken in the limit $\gamma \rightarrow \infty$.

\setcounter{chapter}{3}
\setcounter{equation}{0}
\section*{3 Explicit solutions}

Here we shall display some significant examples of exact solutions to
Eq.(\ref{1.1}) by using the symmetry approach. To this aim, first let us deal
with the generator 
$X^{0}_{4}$
which appears in both the cases $\epsilon=0$ and $\ep\ne 0$. The group
transformations involved by $X_{4}^{0}$ are
\begin{subequations} \lab{X-1}
\begin{eqalignno}
x^{\prime} & = e^{\lambda} x, \lab{X.1a} \\
t^{\prime} & = t, \lab{X.1b} \\
u^{\prime} & = \ga\left[(1+\frac{u}{\ga})e^{\frac{2\lambda}{\ga-2}} 
-1 \right],\lab{X.1c}
\end{eqalignno}
\end{subequations}
where $\lambda$ is the group parameter. A set of basis invariants corresponding
to (\ref{X-1}) is given by
\begin{subequations} \lab{X-2}
\begin{eqalignno}
I_{1} & = (1+\frac{u^{\prime}}{\ga})^{\frac{2-\ga}{2}} 
x^{\prime}\;= \;
(1+\frac{u}{\ga})^{\frac{2-\ga}{2}} x, \lab{X.2a} \\
I_{2} & =t^{\prime}\,= t. \lab{X.2b}
\end{eqalignno}
\end{subequations}
Now, by putting $\rho(t)=I_{1}$, from Eq.(\ref{1.1}) we obtain the reduced
equation 
\be
\rho^{2}_{t}\, +\, \frac{2-\ga}{\ga} \rho \rho_{tt}\,+\,
\frac{2-\ga}{\ga}\epsilon \rho \rho_{t}\, = \,\frac{\ga -1}{\ga}.
\lab{X.3}\ee

Through the change of variable $\rho=e^{W}$, Eq.(\ref{X.3}) becomes
\be
W_{tt} \,+\, \epsilon \ W_{t}\, +\, \frac{2}{2-\ga} \ W^{2}_{t}\,=\,
\frac{\ga -1}{2-\ga}
\ e^{-2W},
\lab{X.4}\ee
which is a kind of modified one-dimensional Liouville equation. For $\ep=0$,
Eq.(\ref{X.4}) can be solved exactly. This occurs via the position
\be
W\, =\, \frac{2-\ga}{2}\  \ln \theta,
\lab{X.5}\ee
which transforms Eq.(\ref{X.4}) into
\be
\theta_{tt} \,=\, \frac{2(\ga -1)}{(2- \ga)^{2}} \ \theta^{\ga -1}.
\lab{X.6}\ee

From (\ref{X.6}) one easily finds
\be
\theta_{t}^{2}\,=\,\frac{4(\ga -1)}{(2-\ga)^{2} \ga} \,\theta^{\ga}\, +\, c,
\lab{X.7} \ee
where $c$ is a constant of integration.

Putting
\be
y^{2}\,=\,1\ +\ a \, \theta^{\gamma}
\lab{X.8} \ee
with
\be
a\,=\,\frac{4(\ga -1)}{c\ga(\ga -2)^{2} },
\lab{X.9} \ee
Eq.(\ref{X.7}) provides
\be
\int dy\,(y^{2}-1)^{\frac{1-\gamma}{\ga}} \,=\,\frac{\ga}{2} \ 
a^{\frac{1}{\ga}} 
\ \sqrt{c}\ t + {\rm const},
\lab{X.10} \ee
with $c > 0$.

In some situations (see Case iv), it may be convenient to write formula (\ref{X.10}) as 
\be
\int({\rm cosh} z)^{-\frac{2}{\ga}} \,dz\,=\,(-1)^{\frac{\ga}{1-\ga}}\, 
\frac{\ga}{2}\  a^{\frac{1}{\ga}}\ 
\sqrt{c} \ t + {\rm const},
\lab{X.11} \ee
which follows from (\ref{X.10}) via the change of variable 
$y= {\rm tanh} z$.

Once the integral at the left-hand side of (\ref{X.10}) has
been calculated, one obtains the {\em exact} solution
\be
u\,=\,\gamma\left[x^{\frac{2}{\ga-2}}\left(\frac{y^{2}-1}{a}
\right)^{\frac{1}{\ga}} -1 \right]
\lab{X.12}\ee
to Eq.(\ref{1.1}) ($\ep \ = \ 0$) 
[see (\ref{X.2a}), (\ref{X.5}), and (\ref{X.8})].

At this point, by way of example, we would like to deal with some particular
value of the parameter $\ga$ which is compatible with an {\em explicit}
expression of $y$ in terms of $t$.

In what follows we shall deal with the cases $i)\,\gamma = \frac{1}{2}$,
$ \;ii)\,\gamma = 3$, and $\; iii)\, \gamma = -{2}.$

The last two choices correspond, respectively, to the Fermi-Pasta-Ulam 
potential\cite{FPU}, and to a potential whose nonlinear part, 
$1/(1-\frac{u}{2})^{2}$, mimics the inverse square potential appearing
in the treatment of the scattering states in conformally invariant Quantum
Mechanics \cite{Sodano}.

{\bf Case i)} For $\gamma = \frac{1}{2}$, from (\ref{X.10}) we have
\be
\int(y^{2}-1) dy\,=\,\frac{1}{4} a^{2} 
\sqrt{c}t ,
\lab{X.111} \ee
with $a=-\frac{16}{9c}$, where the constant at the 
right-hand side of (\ref{X.10})
has been put equal to zero.

Eq.(\ref{X.111}) yields the cubic equation
\be
y^{3} -3y -\frac{3}{4}a^{2} \sqrt{c}t = 0, 
\lab{X.121} \ee
which affords the real solution (\cite{Abramowitz}, p. 17)
\be
y\,=\,s_{1}+s_{2}
\lab{X.131} \ee
where 
\begin{subequations} \lab{X-13}
\begin{eqalignno}\lab{X-131}
s_{1} & = \left(\frac{3}{8}a^{2} \sqrt{c}t
+\sqrt{\frac{9}{64}a^{4}c t^{2} - 1}\right)^{\frac{1}{3}},
\lab{X.131a} \\
s_{2} & =  \left(\frac{3}{8}a^{2} \sqrt{c}t
-\sqrt{\frac{9}{64}a^{4}c t^{2} - 1}\right)^{\frac{1}{3}},
\lab{X.131b}
\end{eqalignno}
\end{subequations}
for $t>\frac{8}{3a^{2}\sqrt{c}}$.

The remaining solutions of (\ref{X.121}) are complex conjugate
functions, and are given
by
\be
y_{1}\,=\,-\frac{1}{2}(s_{1}+s_{2}) + i\frac{\sqrt{3}}{2}(s_{1}-s_{2}), 
\quad \quad 
y_{2}=y_{1}^{*}.
\lab{X.1311} \ee

Now, by inserting (\ref{X.1311}) in (\ref{X.121}) for $\ga=\frac{1}{2}$ we find
the exact solution
\be
u\,=\,\frac{1}{2}\left[\frac{1}{a^{2}}\ x^{-\frac{4}{3}}\ 
(s_{1}^{2}+s_{2}^{2} +
 1) \,-\,1\right].
\lab{X.132}\ee

{\bf Case ii)} For $\ga=3$, Eq.(\ref{X.7}) provides
\be
\theta_{t}^{2}\,=\,\frac{8}{3}\ \theta^{3} \ +\ c,
\lab{X.14} \ee
which can be written as
\be
\theta_{\tau}^{2}\,=\, 4\theta^{3} \ -\ g_{3},
\lab{X.15} \ee
via the rescaling $\tau=\sqrt{\frac{2}{3}}\ t$, with $g_{3}= -c/2$.
Equation (\ref{X.15}) is a special version of the equation
\be
\theta_{\tau}^{2}\,=\,4 \theta^{3}\, -\, g_{2} \theta \,-\, g_{3},
\lab{X.16} \ee 
which is satisfied by the Weierstrass elliptic function 
${\cal{P}}(\tau;g_{2},g_{3})$, where $g_{2}$ and $g_{3}$ are the invariants of 
$\cal P$ (see \cite{Abramowitz}, p. 918).

Concerning Eq.(\ref{X.15}) we shall distinguish two cases, $g_{3}=0$ and 
$g_{3}\ne 0$. For $g_{3}=0$, we have
\be
\theta({\tau})\,=\,{\cal P}(\tau;0,0)\,=\,\frac{1}{\tau^{2}},
\lab{X.17} \ee
which coincides with the first term of the series representation of 
${\cal P}(\tau;g_{2},g_{3})$.

Then, keeping in mind (\ref{X.8}) and (\ref{X.121}), we find
\be
u\,=\,3\left(\frac{3}{2}\ \frac{x^{2}}{t^{2}}
\,-\,1\right).
\lab{X.18} \ee
For $g_{3}\ne 0$, Eq.(\ref{X.15}) gives
\be
\int^{\theta}_{\infty}
\frac{d\al}{\sqrt{4\al^{3}\,-\,g_{3}}}\,=\,
\tau,
\lab{X.19} \ee
that is to say
\be
\theta\,=\,{\cal P}(\tau;0,g_{3}).
\lab{X.20} \ee
Hence, from (\ref{X.12}) we deduce
 \be
u\,=\,3\left(x^{2}\ \theta-1\right)=3\left[x^{2}\ {\cal P}(\tau;0,g_{3})\,-\,1
\right].
\lab{X.21} \ee
We notice that for $c=-2\;\;(g_{3}=1)$, the Weierstrass 
function in (\ref{X.21}) reduces to the equianharmonic case 
${\cal P}(\tau;0,1)$ (\cite{Abramowitz}, p. 652). 

{\bf Case iii)} If $\ga = -2$, from (\ref{X.10}) we easily obtain
\be
u\,=\,-2 \left( \sqrt{\frac{ct^{2}\ -\ a}{x}}\,-\,1
\right),
\lab{X.22} \ee
with $a= \frac{3}{8c}$.

{\bf Case iv)} For $\ga = -1$, the potential (\ref{1.2}) takes 
the form $\phi \sim (1-u)^{-1} - (1+u) $, which resembles a 
special case of the generalized Killingbeck potential
$V=-\frac{A}{r}+Br+Cr^2$ ~\cite{Kill}. We remind the reader that
a Coulomb potential perturbed by a linear term describes the 
spherical Stark effect in hydrogen ~\cite{Letov}.
Putting in (3.11) $\ga=-1$ and $c=-|c|$, we arrive at the 
relation
\be
2z+\sinh 2z = \frac{9}{4}|c|^{\frac{3}{2}}t+const.
\label{Killi} \ee
Curiously enough, by replacing formally the hyperbolic sine by
the exponential, Eq.(\ref{Killi}) becomes an equation of the
Schroeder type, which appears in the bootstrap model and in
renormalization theory, whose analytical structure was investigated
by Hagedorn and Rafelsky \cite{Hagedorn}. These authors obtained 
a solution of the Schroeder equation both as a power series 
expansion and as an integral representation. Thus, it should be
of interest to try to adopt the same strategy in the study
of Eq.(\ref{Killi}).

 
Another nontrivial example of exact solution to Eq.(1.1) can be determined
starting from the linear combination of $X_{3} \ \equiv \ X_{3}^{0}\  =\
\partial_{x}$ and 
$X_{2}\ \equiv \ X_{2}^{0}\ = \ \partial_{t}$: 
\be
V\,=\, v \partial_{x} + \partial_{t},
\lab{Z.15}\ee 
where $v$ is a (real) constant. The group transformations are $x^{\prime}= x + v
\lambda, \quad t^{\prime}= t + \lambda, $ which provide the invariant 
$ \xi = x^{\prime} - v t^{\prime} = x - vt$. Thus, by inserting the variable 
$\xi$ into Eq.(1.1), we get the ordinary differential equation
\be
v^{2}u_{\xi\xi} + \ep v u_{\xi} = \left[\left(1+\frac{u}{\ga}\right)^{\ga -1}
\right]_{\xi\xi},
\lab{Z.16}\ee
which gives
\be
v^{2}u_{\xi} - \ep v u  = \left[\left(1+\frac{u}{\ga}\right)^{\ga -1}
\right]_{\xi} + c_{0},
\lab{Z.17}\ee
where $c_{0}$ is a constant of integration. By choosing $c_{0} = 0$, 
$v^{2} = \frac{\ga -1}{\ga}$ and limiting ourselves to consider those values of 
$\ga$ such that $\ga = -|\ga|$, Eq.(\ref{Z.17}) yields
\be
\left[1- \left(1-\frac{u}{|\ga|}
\right)^{-|\ga| -2}
\right]_{\xi} = \frac{\ep}{v} u,
\lab{Z.18}\ee
from which
\be
\int_{0}^{u} \,\frac{1}{u^{\prime}}
\left[1- \left(1+ \frac{u^{\prime}}{|\ga|}
\right)^{-|\ga| -2}
\right] du^{\prime} = \frac{\ep}{v} (\xi - \xi_{0}),
\lab{Z.19}\ee
$\xi_{0}$ being a constant of integration.

We notice that for $|\ga| \rightarrow \infty$, Eq.(\ref{Z.19}) becomes
\be
\int_{0}^{u} \,\frac{1 - e^{-u^{\prime}}}{u^{\prime}} \;du^{\prime}\,\equiv\,
E{\small in}(u) = 
 \frac{\ep}{v} (\xi - \xi_{0}),
\lab{Z.20}\ee
where $E{\small in}(u)$ denotes the exponential-integral function 
(\cite{Tricomi}, p. 255)
\be
E{\small in}(u) = - E{\small i}(-u) + \ln u + C,
\lab{Z.21}\ee
$E{\small i}(-u) = - \int^{\infty}_{u} 
\frac{e^{-u^{\prime}}}{u^{\prime}} \, du^{\prime}$, 
and $C$ is the Euler-Mascheroni constant,
defined by \cite{Tricomi}
\be
C = - \psi(1) = - \int^{\infty}_{0}\ e^{-t} \ \ln t \, dt,
\lab{Z.22}\ee
where $
\psi(z) = \frac{\Gamma^{\prime}(z)}{\Gamma(z)}$ is the psi (or digamma) 
function
(\cite{Abramowitz}, p. 258).

At this point, let us introduce the function
\bea
\hat{\Gamma}(\alpha, u; |\ga|)\,& = &
\int_{u}^{\infty} \,\left(1 + \frac{u^{\prime}}{|\ga|}\right)^{-(2+|\ga|)} 
\,(u^{\prime})^{\alpha -1} du^{\prime}\,= \nonumber \\
& =&  \hat{\Gamma}(\alpha, 0; |\ga|)\,-\,
\int_{0}^{u} \,\left(1 + \frac{u^{\prime}}{|\ga|}\right)^{-(2+|\ga|)} 
\,(u^{\prime})^{\alpha -1}\;du^{\prime}
\lab{Z.23}\eea
which holds for $0<\al< 2+|\ga|$.

We point out that for $|\ga| \rightarrow \infty$, 
$ \hat{\Gamma}(\alpha, u; |\ga|)$ reproduces the incomplete gamma-function
$\Ga(\al, x)$ (\cite{Abramowitz} p. 260).

Now, since the integral at the right-hand site of (\ref{Z.23}) 
can be written as \cite{Tricomi}
\bea
\int_{0}^{u} \,(u^{\prime})^{\alpha -1}\,
\left(1 + \frac{u^{\prime}}{|\ga|}\right)^{-(2+|\ga|)}
\,du^{\prime}\,  &=&
 \frac{u^{\al}}{\al}\,\ {}_{2}F_{1}(2+|\ga|, \al; \al+1, 
- \frac{u}{|\ga|})\,  =\nonumber \\
& =&  \frac{u^{\al}}{\Ga(2+|\ga|)}\;\sum^{\infty}_{n = 0}\,
\frac{\Ga(2+|\ga|+n)}{(\al +
n)n!} (-1)^{n}\left(\frac{u^{\prime}}{|\ga|}
\right)^{n}. \nonumber\\
\lab{Z.24}\eea
Eq.(\ref{Z.23}) takes the form
\bea
 \hat{\Gamma}(\alpha, u; |\ga|)& =& {\hat \Ga}(\al, 0, |\ga|) -
\frac{u^{\al}}{\al} -\nonumber\\
&& \frac{u^{\al}}{\Ga(2+|\ga|)}
\,\sum^{\infty}_{n=1}\,\frac{\Ga(2+|\ga|+n)}{(\al +
n)n!} (-1)^{n}\left(\frac{u}{|\ga|}
\right)^{n},
\lab{Z.25}\eea
where $\Ga(\cdot)$ is the gamma-function and $_{2}F_{1}$ is the the Gauss
hypergeometric series, respectively (\cite{Abramowitz}, p. 556).

We have (see \cite{Tricomi}, p. 209)
\bea
&&{\rm lim}_{\al \ra 0}
 \hat{\Gamma}(\alpha, u; |\ga|)  =  \nonumber \\
&&=   {\rm lim}_{\al \ra 0}\left[
\frac{\al {\hat \Ga}(\al, 0; |\ga|) -
u^{\al}}{\al} - \frac{u^{\al}}{\Ga(2+|\ga|)}
\sum^{\infty}_{n=1}\,\frac{\Ga(2+|\ga|+n)}{(\al +
n)n!} (-1)^{n}\left(\frac{u}{|\ga|}
\right)^{n} \right] = \nonumber\\
&& =    - \ln u + \ln |\ga| - \psi(2+|\ga|) - C - 
\int_{0}^{u} \frac{1}{u^{\prime}} \left[
\left(1 + \frac{u^{\prime}}{|\ga|}\right)^{-(2+|\ga|)} - 1\right] 
du^{\prime}.
\lab{Z.26}\eea

On the other hand, from (\ref{Z.23}) we obtain (see \cite{Abramowitz} p.255)
\be
{\rm lim}_{\al \ra 0}\;\;
 \hat{\Gamma}(\alpha, u; |\ga|)  = 
\int_{u}^{\infty} \, 
\left(1 + \frac{u^{\prime}}{|\ga|}\right)^{-(2+|\ga|)} \,\frac{1}{u^{\prime}}
du^{\prime}.
\lab{Z.27}\ee
Finally, by comparing (\ref{Z.27}) with (\ref{Z.26}), we find
\bea
\int_{u}^{\infty} \, 
\left(1 + \frac{u^{\prime}}{|\ga|}\right)^{-(2+|\ga|)} \,\frac{1}{u^{\prime}}
du^{\prime}& =& \ln\frac{|\ga|}{u} - \psi(2+|\ga|) - C + \nonumber\\
&&\int_{0}^{u} 
 \frac{1}{u^{\prime}} \left[1\,
-\,\left(1 + \frac{u^{\prime}}{|\ga|}\right)^{-(2+|\ga|)} \right] 
du^{\prime},
\lab{Z.28}\eea
that is (see (\ref{Z.19}))
\bea
 \hat{\Gamma}(0, u; |\ga|) & =& 
\int_{u}^{\infty} \, 
\left(1 + \frac{u^{\prime}}{|\ga|}\right)^{-(2+|\ga|)} \,\frac{1}{u^{\prime}}
du^{\prime} \,=\,\nonumber\\
&=& \frac{\ep}{v}(\xi - \xi_{0}) + 
\ln\frac{|\ga|}{u} - \psi(2+|\ga|) - C . 
\lab{Z.29}\eea

We remark that ${\rm lim}_{|\ga| \ra \infty}\;\;\left[
\ln |\ga| - \psi(2+|\ga|) \right] = 0$ (\cite{Tavole}, p. 945).

Therefore, Eq. (\ref{Z.29}) becomes
\be
{\rm lim}_{|\ga| \ra \infty}\;\;
 \hat{\Gamma}(0, u; |\ga|) \, =\, - E\small i(-u)\, =\,  
\frac{\ep}{v}(\xi\  -\  \xi_{0}) \,-\, \ln u\, -\, C,   
\lab{Z.30}\ee
from which (see (\ref{Z.21}))
\be
 E{\small in}(u) =  
\frac{\ep}{v}(\xi - \xi_{0}).
\lab{Z.31}\ee
This result, corresponding to the continuous Toda case, has been already
obtained in \cite{Ele}. Consequently, the function 
$\hat{\Gamma}(\al, u; |\ga|) $ defined by (\ref{Z.23}) can be considered as an
extended version of the incomplete gamma function
$\;
 {\Gamma}(\al, u)\, =\,  \int_{u}^{\infty} \, e^{-t}\  t^{\al -1} dt
$
(see \cite{Abramowitz} p. 260).

The quantity (\ref{Z.29}), where $\hat{\Gamma}(0, u; |\ga|) $ can be interpreted
as a generalization of the exponential-integral function, constitutes an 
implicit solution of Eq.(1.1).

\setcounter{chapter}{4}
\setcounter{equation}{0}
\section*{4 Linearization and algebraic properties}
Equation (1.1) can be handled within the prolongation scheme 
\cite{Wal}. In 
doing so, let us consider the prolongation equations for Eq.(1.1):
\be
y_{x}^{i}\,=\,F^{i}(u,u_{t};y), \qquad 
y_{t}^{i}\,=\,G^{i}(u,u_{x_{j}};y),
\lab{Z.1}\ee
where $i=1,2,...N$ ($N$ arbitrary), and the set of variables
 $y \equiv \{y^{i}\}$ 
is the pseudopotential \cite{Wal} ($j=1,2,...M$ ($M$ arbitrary)).

The compatibility condition for Eqs. (\ref{Z.1}) gives
\begin{subequations} \lab{Z-2}
\begin{eqalignno}
F^{i}\,= &\frac{\ga}{\ga-1} L^{i}(y)u_{t}\,+\,M^{i}(u,y),
\lab{Z.2a}\\
G^{i}\,=& L^{i}(y)\left(1+\frac{u}{\ga}\right)^{\ga-2} u_{x}\,+\, 
P^{i}(u;y),
\lab{Z.2b}
\end{eqalignno}
\end{subequations}
where $M^{i}\,=\,M^{i}(u;y)$, $P^{i}\,=\,P^{i}(u;y)$ and 
 $L^{i}\,=\,L^{i}(y)$ are defined by
\begin{subequations} \lab{Z-3}
\begin{eqalignno}
\frac{\ga-1}{\ga}\ M_{u}^{i} + [P,L]^{i}\,=\, & \ep L^{i},
\lab{Z.3a}\\
\left(1+\frac{u}{\ga}\right)^{\ga-2}\,[L,M]^{i}\,=\, & P^{i}_{u},
\lab{Z.3b}\\
  {[M,P]}^{i} \,= \,& 0,
\lab{Z.3c}
\end{eqalignno}
\end{subequations}
with $[P,L]^{i}\,=\,P^{k} \frac{\partial L^{i}}{\partial y^{k}}\,-\,
L^{k} \frac{\partial P^{i}}{\partial y^{k}}$, and so on.

In order to explore the prolongation equations (\ref{Z-3}),
 for brevity 
we shall omit the index $i$.

It is noteworthy the following
\begin{prop}
 Let $u$ be a solution of Eq.(\ref{1.5}). Then, the function $y_{2}=
y_{2}(x,t)$ defined by
\be
y_{2x}\,=\,W(u)\, {\rm sinh}\ y_{1}, \qquad
y_{2t}\,=\,S(u)\, {\rm cosh}\ y_{1},
\lab{Z.4}
\ee
\be
y_{1x}\,=\,\frac{\ga}{\ga-1} u_{t}, \qquad
y_{1t}\,=\,\left(1+\frac{u}{\ga}\right)^{\ga-2} u_{x},
\lab{Z.5}
\ee
satisfies the wave equation
\be
y_{2tt}\,=\,\frac{\ga-1}{\ga}\left(1+\frac{u}{\ga}
\right)^{\ga-2} y_{2xx},
\lab{Z.6}\ee
where $W(u)$ and $S(u)$ obey the linear differential equations of the 
Bessel type
\be
W_{uu}\,=\,\frac{\ga}{\ga-1}\left(1+\frac{u}{\ga}
\right)^{\ga-2} W,
\lab{Z.7}\ee
and
\be
S_{uu}\,=\,\frac{\ga}{\ga-1}\left(1+\frac{u}{\ga}
\right)^{\ga-2} S \,+\,
\frac{\ga-2}{\ga}\left(1+\frac{u}{\ga}
\right)^{-1} S_{u}.
\lab{Z.8}\ee
\end{prop}

Equation (\ref{Z.6}) represents a linearized version of Eq.(1.5).

To prove this Proposition, let us look for a solution to Eqs.(\ref{Z-3})
of the form 
\be
M= W(u)V(y), \quad P=S(u)T(y).
\lab{Z.9}\ee
Then Eqs. (\ref{Z-3}) provide
\be
S= \frac{\ga-1}{\ga} W_{u},\qquad S_{u}= \left(1+\frac{u}{\ga}\right)^{\ga-2} 
W,
\lab{Z.10}\ee
and 
\be
{[X_{1},X_{2}]}=X_{3},\quad [X_{1},X_{3}]=X_{2}, \quad [X_{2},X_{3}]=0.
\lab{Z.11}\ee
with $L\ \equiv \ X_{1}$,  $\;V\ \equiv \ X_{2}$,  $\;T\ \equiv \ X_{3}$.

Equations (\ref{Z.10}) imply equations (\ref{Z.7}) and (\ref{Z.8}).
By means of the change of variable

\be
z = 2i \sqrt {\frac{\gamma}{\gamma-1}}
\left(1+\frac{u}{\gamma}\right)^{\frac{\gamma}{2}},
\lab{Z.55}\ee

Eq. (\ref{Z.7}) is transformed into the Bessel equation

\be
z^2 Z_{zz}+zZ_z+\left(z^2-\frac{1}{\gamma^2}\right)Z=0,
\lab{Z.56}\ee

where $Z=Z(z)$ is related to $W$ by

\be
W\,=\,\left(1+ \frac{u}{\ga}\right)^{\half} Z
\left[ 2i \sqrt{\frac{\ga}{\ga-1}} \,\left(1+\frac{u}{\ga}\right)
^{\frac{\ga}{2}}
\right].
\lab{Z.12}\ee

On the other hand, Eq. (\ref{Z.8}) takes the form
\be
z^2 Z_{zz}+z Z_z +
\left[z^2-\left(\frac{\gamma-1}{\gamma}\right)^2\right] Z =0,
\lab{Z.57}\ee

where

\be
S\,=\,\left(1+ \frac{u}{\ga}\right)^{\frac{\ga-1}{2}} 
Z
\left[ 2i \sqrt{\frac{\ga}{\ga-1}}
  \,\left(1+\frac{u}{\ga}\right)^{\frac{\ga}{2}} \right].
\lab{Z.13}\ee

$Z$ stands for a generic Bessel function of index
$\pm \frac{1}{\gamma} $ and $\pm \frac{\gamma-1}{\gamma}$, respectively
(\cite{Abramowitz}, pag. 358).

The commutation rules (\ref{Z.11}) define the Lie 
algebra corresponding to the Euclidean group $E_{2}$ in the plane.
A realization of (\ref{Z.11}) in terms of a two component pseudopotential 
$y\,\equiv\,(y_{1}, y_{2})$ is 
\be X_{1}\,=\, \partial_{y_{1}}, \quad  X_{2}\,=\, 
{\rm sinh}y_{1}\, \partial_{y_{2}}, 
\quad X_{3}={\rm cosh}y_{1}\, \partial_{y_{2}}.
\lab{Z.14}\ee

Therefore, with the help of (\ref{Z.2a}) and (\ref{Z.2b}), Eqs. (\ref{Z.1}) 
take the form expressed by (\ref{Z.4}) and (\ref{Z.5}). Furthermore, by 
differentiating $y_{2x}$ with respect to $x$ and  $y_{2t}$ with respect to 
$t$ (see (\ref{Z.4})) and using (\ref{Z.5}) and (\ref{Z.10}), one arrives at 
the wave equation (\ref{Z.6}).

Finally, we observe that for $\gamma =3 $, i.e. in the case of the
Fermi-Pasta-Ulam potential, Eq. (\ref{Z.7}) is led to the Airy equation
(\cite{Abramowitz}, pag. 446).

\be
W_{\sigma \sigma}=\sigma W,
\lab{Z.59}\ee

via the transformation $\sigma = 2^{-\frac{1}{3}} \left(u+3 \right)$.


The prolongation equations (\ref{Z-3}) offer the possibility of getting
a further insight 
into the algebraic structure of Eq.(1.1).

To this aim, let us write Eqs.(\ref{Z-3}) in the form
\bea
&&\frac{\ga -1}{\ga} P_{z} + [M,L]\,=\, 0,\qquad 
\frac{\ga -1}{\ga} z^{\frac{\ga -2}{\ga -1}}\, M_{z}\,=\, \ep L \ +\ 
[L,P], \nonumber \\
&& [M,P]\, = \,0,
\lab{Z.32}\eea
where $z = \left(1+\frac{u}{\ga}\right)^{\ga -1}$. Then, we look for a 
solution to Eqs.(\ref{Z.32}) such that 
\be
M \,=\, \sum^{\infty}_{k=0} a_{k}(y)z^{k},\qquad
P \,=\, \sum^{\infty}_{k=0} b_{k}(y)z^{k}.
\lab{Z.33}\ee

In the following, we limit ourselves to
 characterize mainly the algebraic structure
of Eq.(1.1) for a particular sequel of values of 
the parameter $\gamma$ (see Proposition 4).
In this case it turns out that a finite-dimensional Lie
algebra is associated with Eq.(1.1).
This algebra is used to write Eq.(1.1) in a potential form, which
allows us to estabilish some analogies between the prolongation 
and the symmetry approaches.
In general, namely for any value of $\gamma$ (provided that $\gamma \ne
0, 1$) a systematic analysis of the algebraic properties of Eq.(1.1),
 which is based on the ansatz (5.2), requires further efforts. 

Substitution from (\ref{Z.33}) into Eq.(\ref{Z.32}) gives the following 
constraints between the coefficients $a_{k}(y)$ and $b_{k}(y)$:
\begin{subequations}\lab{Z.34}
\beq
& k\ \frac{\ga -1}{\ga} b_{k}\  +\  [a_{k-1},L]\,=\, 0,\lab{Z.34-a}\\ 
& \frac{\ga -1}{\ga} (a_{1}+ 2 a_{2}z + 3 a_{3}z^{2} + 4a_{4} z^{3} + \ldots) \
z^{\frac{\ga -2}{\ga -1}} = \nonumber \\
& = \ep L + [L, b_{0}+ b_{1}z + b_{2}z^{2} + b_{3}z^{3} + \ldots], \lab{Z.34-b} 
\eeq
and 
\beq
&[a_{0}, b_{0}] = 0, \nonumber\\
&[a_{0}, b_{1}] + [a_{1}, b_{0}] = 0,   \nonumber\\
&[a_{0}, b_{2}] + [a_{1}, b_{1}] + [a_{2}, b_{0}] = 0, \nonumber\\
& \ldots\ldots\ldots\ldots\ldots\ldots \nonumber \\
& \sum_{k=1}^{N} [a_{k-1}, b_{N-k}] = 0, \lab{Z.34-c} 
\eeq
\end{subequations}
with $N$ arbitrary.

The following property holds
\begin{prop}

If $\ep \ne 0$ and $\nu\,=\,\frac{\ga -2}{\ga -1}$ is such that $ 
\nu \ \ne \ \ldots, -3\, , -2\, , -1\, , 0\, , 1\, , 2\, , 3\, \ldots,$ Eqs.(\ref{Z.34}) imply the
finite-dimensional Lie algebra $\cal L$
\bes \lab{Z.35}
\beq
{[a_{0}, b_{0}]}\,=\, & \,[a_{0}, b_{1}]\,=\,[L, b_{1}]\,=\,0, \lab{Z.35-a}\\
{[b_{0}, b_{1}]}\,=\, & \,\ep b_{1}, \lab{Z.35-b}\\
{[b_{0},L]}\,=\,&  \ep L, \lab{Z.35-c} \\
{[a_{0}, L]}\,=\, & \frac{1}{\nu - 2} b_{1}, \lab{Z.35-d}
\eeq
\ees

for $\ep \ne 0$, and the finite-dimensional Lie algebra ${\cal 
L^{\prime}}$

\bes \lab{Z.44}
\beq
{[a_{0}, b_{0}]}\,=\, & \,[a_{0}, b_{1}]\,=\,[L, b_{1}]\,=\,{[b_{0},
b_{1}]}\,=\,{[b_{0},L]}\,=\,0, \lab{Z.44-a}\\
{[a_{0}, L]}\,=\, & \frac{1}{\nu - 2} b_{1}, \lab{Z.44-b} 
\eeq
\ees

for $\ep = 0$.
\end{prop}

The proof is straightforward. In fact, under the assumption that $\nu \
\ne \ldots,$ $ -3, -2, -1, \ldots, $ from (\ref{Z.34-b}) we obtain $a_{k}\ =\
  0$, $\;[b_{k}, L]\ =\ 0 \quad (k = 1,2,3, \ldots )$, and the
  commutation relation (\ref{Z.35-c}). Furthermore, Eq.(\ref{Z.34-c})
  entails the commutation relation (\ref{Z.35-d}) and $b_{2}\ =\ b_{3}\
  = \ldots \ = 0$. Furthermore, Eq.(\ref{Z.35-b}) can be determined by elaborating Eq.(\ref{Z.35-c}) via the Jacobi
    identity applied to $[a_{0}, [b_{0},L]]$.
The commutation relations (\ref{Z.44}) emerge immediately.

A matrix representation of $\cal L$ is 

\bea
& a_{0}\, =\,  \pmatrix{0 & -1 & 0\cr
 0 & 0 & 0 \cr
 0 & 0 & 0 \cr},
 \qquad \qquad &  b_{0} \, =\, \pmatrix{\ep & 0 & 0\cr
 0 & \ep
    & 0 \cr 
0 & 0 & 0\cr}, 
\nonumber\\
& b_{1}\, =\, \pmatrix{0 & 0 & 1\cr
 0 & 0 & 0 \cr
 0 & 0 & 0 \cr},
\qquad \qquad &  L \, =\, \pmatrix{ 0 & 0 & 0\cr
 0 & 0 
 &
    \frac{1}{2- \nu} \cr 0 & 0 & 0 
\cr}.
\eea

Then, Eqs. (\ref{Z.1}) take the form

\bes\lab{Z.36}
\beq
& {\left(\begin{array}{l} y_1 \\ y_2 \\ y_3 \end{array}\right)}_x\,
=\, \pmatrix{0 & 0 & 0 \cr
-1& 0 & 0 \cr
0 &u_t& 0 \cr}
\left(\begin{array}{l} y_1 \\ y_2 \\ y_3
\end{array}\right),\lab{Z.36-a}\\
\nonumber\\
& {\left(\begin{array}{l} y_1 \\ y_2 \\ y_3 \end{array}\right)}_t\,
=\, \pmatrix{\epsilon & 0 & 0 \cr
0 &\epsilon & 0\cr
\left(1+\frac{u}{\gamma}\right)^{\gamma -1} & \frac{\gamma
-1}{\gamma}u_x\left(1+\frac{u}{\gamma}\right)^{\gamma -2} &
0\cr}
\left(\begin{array}{l} y_1 \\ y_2 \\ y_3
\end{array}\right),\lab{Z.36-b}
\eeq
\ees
\par
\par from which $y_1=\lambda_1 e^{\epsilon t} $, $y_2=\left(\lambda_0
-\lambda_1 x \right)e^{\epsilon t} $, and 

\bes\lab{Z.37}
\beq
&{{y_3}_\zeta}=-\frac{1}{\lambda_1}e^{\epsilon t} u_t \zeta ,\lab{Z.37-a}\\
&{{y_3}_t}=-\lambda_1e^{\epsilon t}\zeta^2
 \frac{\partial}{\partial\zeta}\frac{\left(1+\frac{u}{\gamma}\right)^{\gamma -1}}
{\zeta},\lab{Z.37-b}
\eeq
\ees

where $\zeta = \lambda_0 -\lambda_1 x$, and $\lambda_0 $, $\lambda_1 $ are
constants of integration. Here $y_3$ can be interpreted as a 
potential variable.

Equations (\ref{Z.36-a}) and (\ref{Z.36-b})  allow us to find, in
theory, special explicit solutions to
Eq.(1.1). For example, let us assume that
$u_t=\beta (t)\zeta^{\frac{2}{\gamma -2}}$. Then, Eqs.(\ref{Z.37-a}) and
(\ref{Z.37-b}) give rise to the differential equation for $\beta (t)$:

\be\lab{Z.38}
\beta = \gamma b^{\frac{1}{\gamma -1}}\frac{d}{dt} \left(\dot \beta +
\epsilon \beta\right)^{\frac{1}{\gamma -1}}
\ee

where $b=\frac{\left(\gamma -2 \right)^2}{2\gamma \left(\gamma
-1\right)}{\lambda_1}^2 $. 

Equation (\ref{Z.38}) can be written as 

\be\lab{Z.39}
\ddot V+\epsilon \dot V =k V^{\gamma -1},
\ee

($k=\frac{b^{-\frac{1}{\gamma-1}}}{\gamma}$) through the position

\be
\dot \beta +\epsilon \beta = V^{\gamma -1}.
\ee

For $\ep = 0$, we may use for the algebra ${\cal  L^{\prime}}$  the same
representation (5.6), where now $b_0$ is the null matrix.
In this case Eq.(\ref{Z.39}) coincides with Eq.(3.6), which leads to
explicit solutions of the original equation (1.5) for special values of
the parameter $\gamma$ (see Sect.3).
We remark that via the change of variable $V=e^{\frac{2}{2-\gamma}W} $
and by a suitable choice of the free constant $k$, Eq.(\ref{Z.39})
becomes Eq.(3.4), which
arises from the symmetry operator ${{X_4}^0}$ in the context of the Lie
group theory. This fact suggests the existence of a possible link
between the prolongation method and the symmetry approach. However, our
result is indicative only, and many aspects of the problem remain to be
elucidated.

\setcounter{chapter}{5}
\setcounter{equation}{0}

\section*{5 Comments}
We have dealt with a nonlinear field equation arising as the 
continuous limit of a lattice model containing many cases of
physical significance (the harmonic, Toda, Fermi-Pasta-Ulam,
Coulomb-like lattices, and others). This equation, which is new
at the best of our knowledge, can be considered as a paradigm
for the application of different analytical procedures. The
addition of a perturbative friction-like term has allowed us
to check the limits of validity of these methods. These are the
symmetry and approximate symmetry approach, and the prolongation
technique. The joint resort to these methods leads to the 
discovery of some interesting properties of Eq.(\ref{1.1}),
which have been expounded in the Introduction.

The spirit of this paper is both of methodological and speculative
character. The results obtained are a challange to apply the same
strategy to investigate a three-dimensional version of Eq.(\ref{1.1}).
This purpose might be important in nonlinear field theories
based on the deformation of the algebra used in the study of
dispersionless field equations \cite{Strachan}.



\end{document}